# Prédiction de l'évolution granulométrique et morphologique d'une poudre dans un four tournant

F. Patisson[a], C. Ablitzer-Thouroude[b], S. Hébrard[a,b], D. Ablitzer[a]

[a] LSG2M, UMR 7584 CNRS-INPL, Ecole des Mines, Parc de Saurupt, CS 14234, 54042 Nancy Cedex
[b] Commissariat à l'Energie Atomique, CEA-Cadarache, DEC/SPUA/LCU, bâtiment 315, 13108 Saint-Paul-lez-Durance,

**RESUME**

La poudre d'$UO_2$ utilisée pour la fabrication des pastilles de combustible nucléaire est obtenue en France par une conversion en voie sèche d'$UF_6$ gazeux. Le procédé comporte deux étapes : hydrolyse en $UO_2F_2$, puis pyrohydrolyse réductrice en $UO_2$ en four tournant. Les caractéristiques physiques (morphologie, distribution granulométrique) de la poudre d'$UO_2$ obtenue en sortie de four conditionnent ses propriétés d'usage (frittabilité, coulabilité et tenue mécanique à cru). Nous avons développé un modèle décrivant l'évolution morphologique de la poudre dans le four tournant, de façon à disposer d'un outil de prédiction des caractéristiques morphologiques de la poudre d'$UO_2$ en fonction de ses conditions d'élaboration. Une première partie du travail a consisté à modéliser le transport de la poudre dans le four, en décrivant notamment les échanges entre la phase dense (lit de poudre) et la phase dispersée (pluie de particules en suspension). Une des originalités du modèle développé est la prise en compte fine du rôle des releveurs pour le calcul des variables dynamiques. La seconde partie a consisté à identifier, décrire et coupler au modèle dynamique précédant les phénomènes responsables de l'évolution morphologique et granulométrique de la poudre dans le four tournant. On considère une population d'agglomérats fractals dont le nombre et la taille évoluent par agglomération brownienne, agglomération par sédimentation différentielle, préfrittage, fragmentation, et transformations chimiques par ex-nucléation et croissance. Ce modèle emploie le formalisme des bilans de population et la distribution granulométrique est discrétisée en sections. Les résultats des calculs dynamiques et morphologiques sont comparés aux mesures disponibles. Enfin, on analyse l'influence respective des différents mécanismes d'évolution morphologique sur la distribution granulométrique finale.

**MOTS-CLES :** poudre, dioxyde d'uranium, évolution morphologique, granulométrie, four tournant

## INTRODUCTION

La poudre de dioxyde d'uranium $UO_2$ utilisée pour la fabrication de pastilles de combustible nucléaire peut être produite par le procédé de conversion en voie sèche d'$UF_6$ gazeux. Ce procédé réalise successivement une hydrolyse de l'$UF_6$ gazeux en poudre de difluorure d'uranyle $UO_2F_2$, puis une pyrohydrolyse réductrice de la poudre d'$UO_2F_2$ en poudre d'$UO_2$. La première étape est réalisée dans un réacteur à jets. La seconde se déroule dans un four tournant accolé au premier réacteur.

Certaines des caractéristiques physiques de la poudre produite, comme la taille et la forme des particules et des agrégats ou agglomérats qui la constituent, conditionnent ses propriétés d'usage (frittabilité, coulabilité, tenue mécanique à cru, etc.). Nous emploierons ici le terme morphologie pour désigner globalement la structure de la poudre, en particulier les distributions de tailles et les formes des particules primaires et des agglomérats. Dans cet article, nous présentons un travail de modélisation de l'évolution morphologique de la poudre dans le four tournant qui a été réalisé dans le but d'améliorer notre connaissance de l'influence des paramètres de cette étape du procédé sur les caractéristiques morphologiques du produit final. La Figure 1, qui présente des poudres d'$UO_2F_2$ et d'$UO_2$ typiques, illustre l'étendue de cette évolution morphologique.

La description de l'évolution morphologique de la poudre au cours de sa transformation dans le four tournant nécessite d'abord de décrire la dynamique de la phase solide dans le four. Cet aspect est l'objet de la première partie de cet article. Dans la deuxième partie, nous détaillons les principaux phénomènes physico-chimiques responsables d'une modification de la morphologie et nous présentons le modèle développé pour en rendre compte, ainsi que quelques-uns de ses résultats.

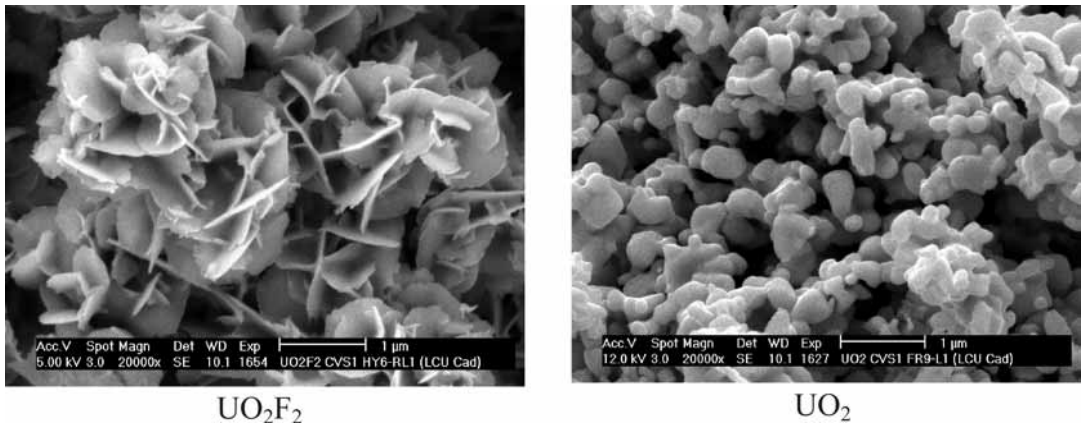

Figure 1 – Micrographies MEB des poudres initiale et finale

**ASPECTS DYNAMIQUES**

Le four tournant du réacteur de pyrohydrolyse est pourvu d'équipements internes (chicanes et cornières) destinés à améliorer les échanges de matière entre gaz et solides. Le mouvement de la poudre dans un tube tournant muni de tels dispositifs diffère donc, et est plus complexe, de celui rencontré classiquement dans les tubes qui en sont dépourvus.

Les fours tournants inclinés dépourvus d'équipements internes sont employés dans nombre de procédés de transformation de la matière (ex. : séchage, calcination, pyrolyse) et fonctionnent le plus souvent dans le mode dit de roulement pour ce qui est de l'écoulement transversal. Les particules en surface ("couche active") dévalent le talus du lit en suivant la ligne de plus grande pente, s'arrêtent et sont réintégrées aléatoirement au sein du lit qui tourne solidairement avec la paroi. L'avance axiale du solide n'a lieu que sous l'effet de l'inclinaison du tube et lors du roulement en surface. Une particule au sein du lit n'a qu'un mouvement circulaire sans composante axiale. La trajectoire qui en résulte est schématisée sur la Figure 2.

Dans le cas où le tube est pourvu de releveurs, ceux-ci se chargent de poudre en piégeant celle située au fond du tube et en réceptionnant la poudre déversée par les releveurs situés au-dessus (Figure 3). A partir d'une position angulaire notée $\theta_{init}$, un releveur commence à décharger la poudre qu'il contient. Les particules déchargées à partir d'un releveur chutent sous l'effet de la gravité et sont en outre entraînées par le courant gazeux. Dans un tel système, il est utile de distinguer la phase solide dispersée (pluie chutant des releveurs) et la phase solide "dense" (poudre en lit au fond du tube ou piégée dans les releveurs). Le calcul de la rétention solide et du débit axial doit tenir compte de ces deux phases. L'avance axiale du solide en phase dense résulte, comme dans les tubes vides, de l'inclinaison du tube par rapport à l'horizontale et de l'inclinaison locale du lit par rapport à l'axe du tube, tandis que celle en phase dispersée résulte de l'inclinaison du tube et de l'entraînement des particules par le gaz lors de leur chute.

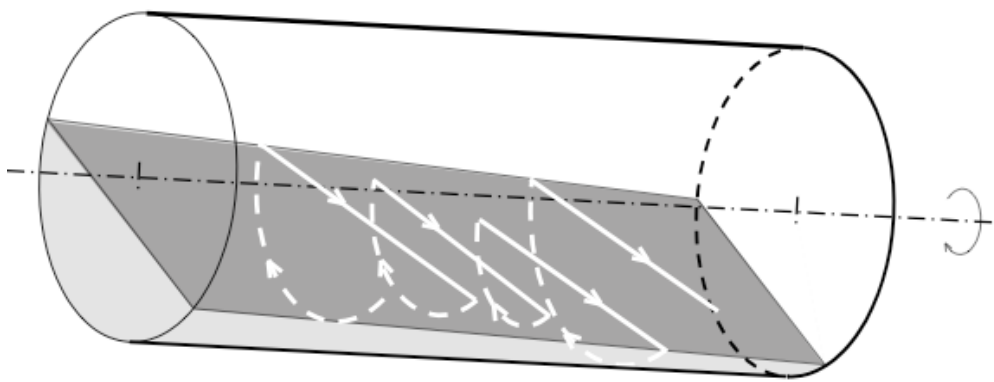

Figure 2 – Trajectoire typique d'un grain de solide dans un four tournant sans releveurs

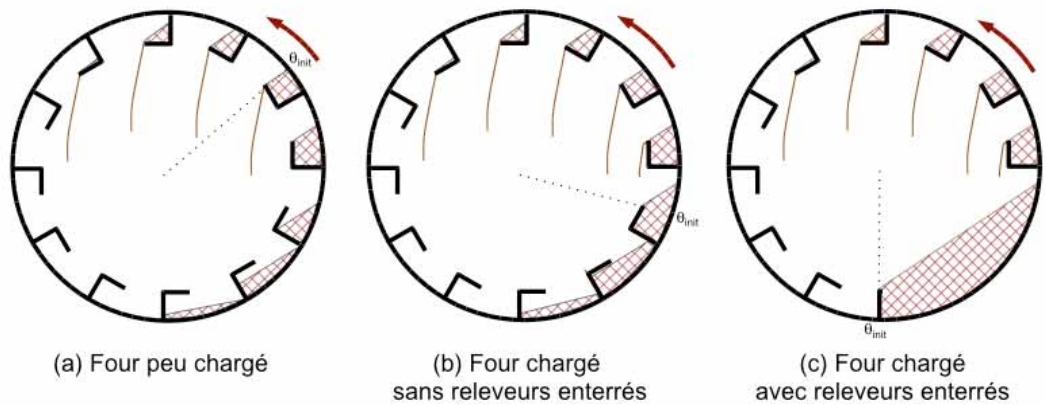

(a) Four peu chargé  (b) Four chargé sans releveurs enterrés  (c) Four chargé avec releveurs enterrés

Figure 3 – Section transversale d'un tube tournant muni de releveurs, pour différents degrés de chargement

Le degré de chargement du four, c'est-à-dire la fraction de volume disponible occupée par la poudre, en partie illustré par la figure ci-dessus, est un paramètre crucial, qui influe à la fois sur l'écoulement radial et axial et sur la répartition de la poudre entre phases dense et dispersée. Les autres grandeurs dynamiques susceptibles d'influencer l'évolution morphologique sont les flux massiques et les rétentions de chacune des phases, ainsi que le temps de séjour moyen dans le tube.

Nous avons donc été conduits à développer un modèle dynamique capable de calculer l'ensemble de ces grandeurs. Nous nous sommes pour cela grandement inspirés du modèle de Sherritt et Caple [1] développé à l'origine pour simuler l'écoulement de copeaux de bois dans des fours de séchage. Il s'agit d'un modèle de type géométrique qui détermine la plupart des variables dynamiques à partir du calcul de la vitesse de déchargement d'un releveur. Nous avons cependant modifié ce modèle pour assurer la continuité du calcul du taux de rétention de la phase dense entre les cas avec et sans releveurs, ainsi que pour pouvoir tenir compte de différents angles de talus.

La Figure 4 présente deux profils de remplissage calculés pour deux vitesses de rotation du four par le modèle dynamique, ainsi que les profils correspondants obtenus grâce à la relation de Kramers et Croockewit [2], relation classiquement utilisée pour les fours dépourvus de releveurs. Ces résultats illustrent l'influence des releveurs et montrent que, suivant les conditions opératoires, les corrélations utilisées pour les fours sans releveurs peuvent sur- ou sous-estimer le taux de chargement d'un four avec releveurs.

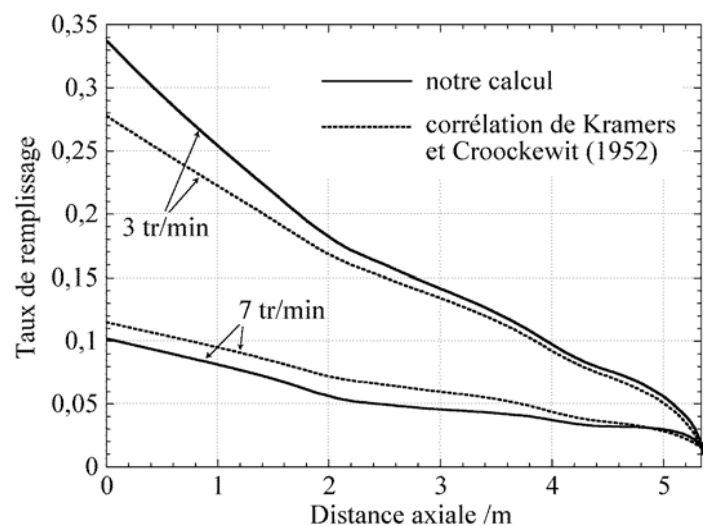

Figure 4 – Profils de degrés de chargement calculés pour deux vitesses de rotation du four

**MODELE D'EVOLUTION MORPHOLOGIQUE**

Pour décrire l'évolution morphologique de la poudre lors de son transport et de ses transformations dans le four tournant, nous avons dû définir une représentation géométrique satisfaisante de la poudre, déterminer les mécanismes responsables de l'évolution morphologique et quantifier leur cinétique, modéliser mathématiquement les phénomènes correspondants et, enfin, résoudre numériquement les équations obtenues. Ces différents aspects de la modélisation sont décrits ci-dessous.

**Description des particules**

On considère que la poudre est composée d'agglomérats fractals de particules dites primaires (Figure 5). La taille des agglomérats et la taille des particules primaires varient le long du four. On tient compte d'une distribution de tailles des agglomérats, qui représente donc la distribution granulométrique de la poudre, tandis que les particules primaires sont toutes supposées, à une abscisse du four donnée, sphériques et de même taille. Les premières grandeurs caractéristiques d'une particule de poudre, i.e. un agglomérat, sont donc le nombre de particules primaires, la taille des particulaires primaires et la dimension fractale de l'agglomérat. On peut en déduire le volume de l'agglomérat, sa surface, son diamètre de collision, sa porosité, etc.

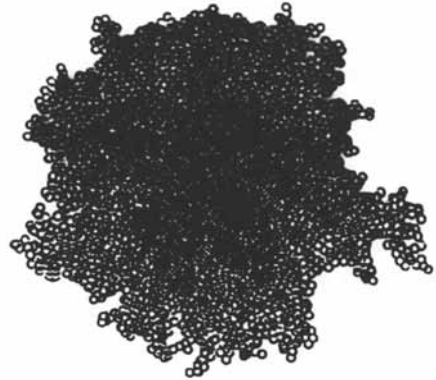

Figure 5 – Représentation schématique d'une particule de poudre comme un aggloméré fractal de particules primaires

**Les mécanismes d'évolution morphologique**

Lors de son avancée dans le four tournant, la poudre évolue du fait d'un ensemble de transformations que nous avons distinguées suivant leur nature : physique, thermique ou chimique. Une analyse de la littérature et la connaissance des conditions locales dans le four nous ont permis d'identifier a priori les phénomènes suivants comme pouvant jouer un rôle dans l'évolution morphologique. Ainsi, l'agglomération et la fragmentation des particules sont les principaux mécanismes d'évolution de nature physique. Le frittage, phénomène activé par les hautes températures rencontrées, engendre une réduction de surface des agglomérats. Enfin, les transformations chimiques, défluoration et réduction, en faisant disparaître et apparaître de nouvelles phases, modifient la forme et la taille des particules primaires et par conséquent celle des agglomérats.

L'agglomération peut concerner les particules de la phase dispersée et celles de la phase dense. Si l'agglomération en phase dense est un phénomène mal connu, celle en phase dispersée a été largement étudiée. Pour cette dernière et dans les conditions aérodynamiques et thermiques du four tournant, plusieurs contributions s'ajoutent : l'agglomération brownienne, liée au mouvement brownien des molécules de gaz, et l'agglomération par sédimentation différentielle, qui résulte des différences de vitesse de chute (à partir des releveurs) entre gros et petits agglomérats. Nous avons montré que l'agglomération turbulente était négligeable. Dans la phase dense, nous avons estimé que les particules pouvaient s'agglomérer sous l'effet de gradients de vitesse lors de l'écoulement dans la couche active et nous avons conçu une représentation du phénomène par analogie avec l'agglomération par cisaillement dans les gaz. Enfin, en ce qui concerne le phénomène de fragmentation, qui s'oppose à et réduit les effets de l'agglomération, nous avons supposé une fragmentation dite homogène, c'est-à-dire plus fréquente pour les gros agglomérats, et donnant des fragments de volumes égaux.

Le frittage des particules est un phénomène relativement complexe, mais bien connu et souvent décrit, notamment dans le domaine de la métallurgie des poudres. Il conduit, par formation de ponts entre particules, à une diminution de la surface spécifique, puis à la diminution du volume des pores et, dans ses stades ultimes, à la densification, c'est-à-dire l'élimination de la porosité fermée. Dans le four tournant, il s'agit plutôt d'un préfrittage (par opposition au frittage ultérieur de l'$UO_2$) qui affecte les particules primaires dans les agglomérats et demeure d'un avancement très limité. Malgré cela, son importance est certainement cruciale pour la cohésion des agglomérats.

La conversion chimique d'$UO_2F_2$ en $UO_2$ fait appel à de nombreuses étapes qui font apparaître des intermédiaires réactionnels. Suivant les conditions expérimentales (température, concentrations en $H_2$ et $H_2O$), on peut notamment former les oxydes $UO_3$, $UO_{2,9}$ et $U_3O_8$ [3-5]. Lors de chacune de ces réactions gaz-solide, la morphologie des grains change du fait de la germination et de la croissance d'une nouvelle phase solide (le produit) en remplacement d'une autre (le réactif). Pour notre modélisation, nous avons supposé que, pour la plupart des réactions, cela se traduisait par une simple variation du volume des particules primaires, qui peut être localement calculée connaissant les volumes molaires des différentes phases et supposant connue la composition. En revanche, en ce qui concerne la première étape, la défluoration d'$UO_2F_2$ en $UO_{2,9}$, Chaillot [3] a remarqué qu'elle faisait intervenir un phénomène d'ex-nucléation. Les germes d'$UO_{2,9}$ qui apparaissent à la surface des grains d'$UO_2F_2$ commencent à croître, puis sont éjectés (ex-nucléés) quand ils ont atteint une taille critique d'environ 10 nm, sans doute sous l'effet de contraintes mécaniques. Il s'ensuit que la population des nouvelles particules n'a plus aucune filiation morphologique avec la phase mère. Nous en avons tenu compte en considérant dans notre modèle deux populations d'agglomérats : celle des particules initiales d'$UO_2F_2$, qui disparaît rapidement, et celle des particules d'$UO_x$ (avec x passant de 3 à 2). La Figure 6 synthétise ces phénomènes d'évolution morphologique d'origine chimique.

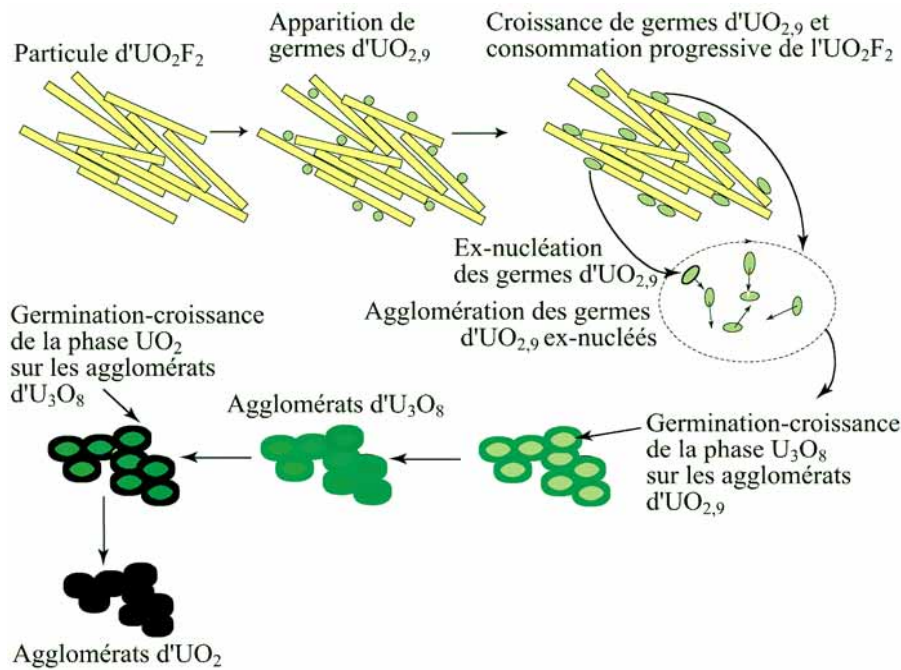

Figure 6 – Evolution morphologique d'origine chimique

**Bilans de population**

Le formalisme des bilans de population est utilisé pour modéliser l'ensemble des phénomènes précédents et accéder à l'évolution en nombre et en taille des agglomérats le long du four. La distribution granulométrique est discrétisée en sections en utilisant la méthode de Kumar et Ramkrishna [6-8] qui présente sur d'autres l'avantage d'assurer à la fois la conservation du nombre et du volume total des particules. La technique du pivot mobile est appliquée pour traiter des termes de croissance chimique. Trois grandeurs font l'objet de ce type de bilan : le nombre, le volume et la surface des agglomérats. La méthode numérique a été pour partie validée en comparant avec succès l'évolution granulométrique prédite par le modèle à une solution analytique de la littérature [9] dans le cas d'une simple agglomération à noyau constant.

Ces bilans sont couplés à la partie dynamique tout d'abord en considérant une évolution spatiale (le long de l'axe du four) et non temporelle (comme dans les modèles usuels d'agglomération). On introduit ainsi les flux massiques des phases dense et dispersée. On tient également compte du fait que certains termes sources, notamment d'agglomération et de fragmentation, reflètent des situations dynamiques différentes et doivent donc différer suivant la phase considérée : rideau de poudre chutant d'un releveur, couche active à la surface du talus au fond du four, etc.

**Détermination des paramètres**

De nombreux paramètres physiques, thermiques et cinétiques apparaissent dans le modèle morphologique. Ils ont été déterminés soit à partir d'expériences entreprises à cet effet, soit grâce à des données de la littérature. Les phénomènes et les produits concernés étant très spécifiques, il n'a malheureusement pas été possible de déterminer de façon exhaustive l'ensemble de ces paramètres.

Le noyau d'agglomération brownienne est calculé suivant la relation de Fuchs qui couvre à la fois le régime moléculaire et le régime continu [10]. Pour déterminer le noyau d'agglomération par sédimentation différentielle, nous avons modifié le calcul de la vitesse de Stokes en recalculant un diamètre hydraulique qui tienne compte de la nature fractale des agglomérats [11]. L'efficacité de capture de l'agglomération brownienne est classiquement considérée comme égale à 1. Celle de l'agglomération sédimentaire a été peu étudiée, nous l'avons prise égale ou inférieure à 0,1, comme préconisé dans [12, 13]. L'agglomération par cisaillement dans la phase dense n'a pu être prise en compte faute de données. La fréquence de fragmentation et le nombre de fragments sont difficiles à déterminer et demeurent deux paramètres du modèle.

Concernant le frittage, les températures des différents composés dans le four sont inférieures aux températures de fusion, ce qui permet de considérer que les mécanismes de diffusion les plus probables sont la diffusion en surface et la diffusion aux joints de grains. D'après [14], le temps caractéristique de frittage peut s'écrire formellement de la même manière pour ces deux mécanismes, en faisant intervenir une constante cinétique et une énergie d'activation. Nous avons déterminé expérimentalement ces deux paramètres pour les différents composés ($UO_2F_2$, $U_3O_8$ et $UO_2$) à partir d'essais de frittage en fours statiques.

Pour ce qui est des transformations chimiques, nous avons supposé connus les profils de composition des solides et retenu ceux tirés des travaux de Debacq sur le même système [15], qui font intervenir uniquement les espèces $UO_2F_2$, $U_3O_8$ et $UO_2$. On suppose alors que la transformation d'$UO_2F_2$ donne $U_3O_8$ (au lieu d'$UO_{2,9}$) qui apparaît sous forme de nouvelles petites particules correspondant aux germes ex-nucléés. Une taille de germes de 50 nm a été retenue car une étude de sensibilité a montré que les calculs étaient dans ce cas beaucoup plus rapides qu'avec des germes de 10 nm, tout en modifiant à peine le résultat final. Par ailleurs, les variations de volume liées aux phénomènes de croissance (consommation d'$UO_2F_2$ et changement de phase $U_3O_8/UO_2$) sont réparties proportionnellement au volume de chaque classe.

Les autres paramètres nécessaires sont des caractéristiques physicochimiques usuelles des différents oxydes d'uranium.

La distribution granulométrique initiale de l'$UO_2F_2$ est mal connue. Les mesures que nous avions entreprises pour la déterminer n'ont pas été concluantes. Nous l'avons supposée log-normale avec un diamètre moyen de 3 μm et un écart-type relatif de 1,3. Signalons ici que, compte-tenu du phénomène d'ex-nucléation, cette distribution a en fait peu d'influence sur la distribution granulométrique finale de l'$UO_2$. Enfin, la dimension fractale des agglomérats a été déterminée égale à 2,6 à partir de mesures de masse volumique des agglomérats et de taille des agglomérats et des particules primaires sur différents lots d'$UO_2$.

Du point de vue numérique, une discrétisation en 80 classes a été utilisée, avec une raison géométrique de 1,33.

**Quelques résultats**

La Figure 7 montre l'évolution simulée de la distribution des agglomérats d'$UO_x$ en fonction de l'abscisse du four z, en tenant compte de l'ensemble des mécanismes d'évolution morphologique. La distribution à z=0,01 m reflète essentiellement la production de très nombreux petits germes par le phénomène d'ex-nucléation. A z=0,5 m, on note un grossissement de ces petites particules dû à l'agglomération. A z=1 m, la défluoration est terminée, on ne forme plus de germes et leur nombre commence à décroître. A partir de z=1,5 m, ceux-ci ont complètement disparu et la distribution continue d'évoluer, jusqu'en sortie, vers de plus grosses particules. La distribution finale calculée est proche d'une distribution log-normale avec cependant un nombre de fines et de grosses plus important. Le diamètre moyen des agglomérats, de l'ordre de 3 μm, est proche des diamètres observés expérimentalement. Il en va de même pour la dispersion de la distribution. L'étalement de la distribution vers les plus grosses particules est également observé expérimentalement, quoique dans une moindre mesure. La taille des particules primaires calculée (0,44 μm – non représentée sur la figure) semble par contre surestimée par rapport aux tailles que l'on peut déduire des mesures de surface spécifique (0,2 à 0,3 μm).

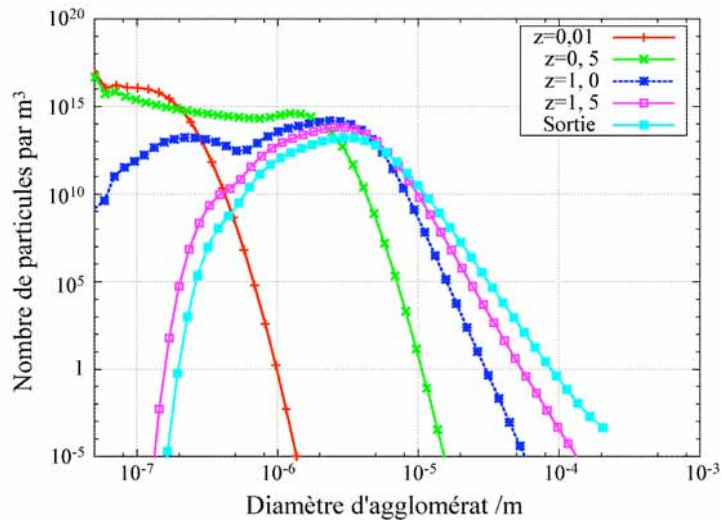

Figure 7 – Evolution simulée de la distribution granulométrique des agglomérats d'UO$_x$ pour l'ensemble des mécanismes d'évolution morphologique

La Figure 8 montre les distributions finales simulées des agglomérats d'UO$_2$ en sortie de four en fonction des mécanismes d'évolution morphologique pris en compte et permet donc de comprendre voire de quantifier l'influence de chacun d'eux. L'introduction du frittage diminue un peu les tailles finales d'agglomérats car il atténue l'agglomération brownienne en diminuant les diamètres de collision des agglomérats. L'agglomération par sédimentation est le mécanisme responsable d'un étalement de la distribution vers les grosses particules. L'effet est très sensible, un facteur d'efficacité de capture trop élevé de cette agglomération conduirait à des résultats irréalistes. Comme attendu, la fragmentation a un effet inverse de réduction des grosses particules.

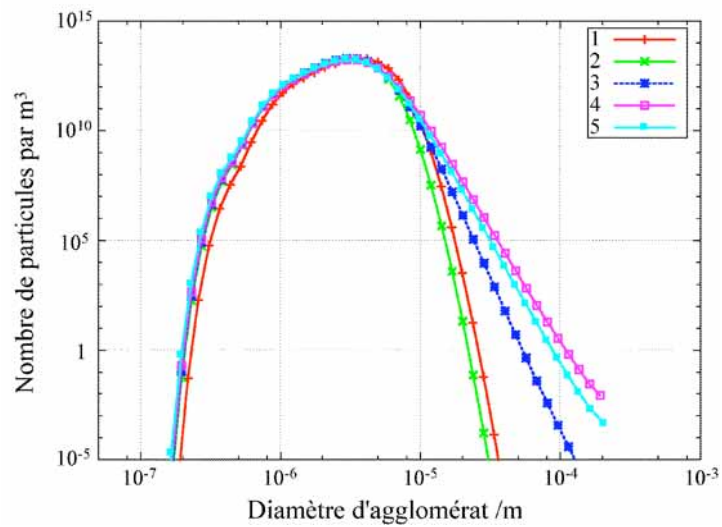

Figure 8 – Distribution granulométrique finale calculée en fonction des mécanismes d'évolution morphologique : 1= agglomération brownienne seule ; 2=1+frittage ; 3=2+agglomération sédimentaire d'efficacité 0,05 ; 4=2+agglomération sédimentaire d'efficacité 0,1 ; 5=4+fragmentation

**CONCLUSION ET PERSPECTIVES**

Le modèle qui a été présenté dans cet article présente l'avantage et l'originalité de coupler étroitement une modélisation de l'écoulement de la poudre dans un four tournant muni d'équipements internes et une modélisation des principaux phénomènes d'évolution morphologique. Il permet de comparer l'influence relative de différents phénomènes sur la morphologie du produit final. Il pourra constituer à terme un outil de prédiction de l'influence des conditions opératoires sur la morphologie.

Le modèle actuel présente cependant quelques insuffisances liées à des simplifications du système. En particulier, il faudrait que le profil de composition soit un résultat et non une donnée du modèle. Il sera pour cela nécessaire de coupler le modèle d'évolution morphologique avec les cinétiques de réaction. L'une des conséquences de ce profil de composition imposé, associé à l'hypothèse d'ex-nucléation de germes lors de la transformation du composé $UO_2F_2$, est que les caractéristiques morphologiques d'$UO_2F_2$ en entrée de four apparaissent sans effet sur la poudre d'$UO_2$. Or, des résultats expérimentaux obtenus avec des poudres d'$UO_2F_2$ de caractéristiques différentes montrent une influence de ces caractéristiques sur la poudre d'$UO_2$ finale. Par ailleurs, plusieurs paramètres restent indéterminés en raison de la difficulté à réaliser des estimations expérimentales (efficacité de capture de l'agglomération par sédimentation différentielle, coefficients de la loi de fragmentation, paramètres des lois de frittage...).

Une validation quantitative du modèle nécessitera de déterminer les caractéristiques de la poudre d'$UO_2$ pour diverses conditions opératoires et de confronter les résultats obtenus à ceux qui sont accessibles par l'expérience (notamment la surface spécifique).